\documentclass[a4paper, 10pt]{article}
\usepackage[left=35mm, right=35mm, top=35mm, bottom=25mm]{geometry}
\usepackage{amsthm,amsmath}
\usepackage[nohyperlinks]{acronym}
\usepackage{amsmath}
\usepackage{hyperref}
\usepackage{caption}
\usepackage{subcaption}
\usepackage{graphicx}
\usepackage{authblk}
\usepackage{float}
\title{Exploiting routinely collected severe case data to monitor and predict influenza outbreaks}

\begin{document}
\author[1]{A. Corbella$ ^{\dagger \S} $}
\author[2]{X.-S. Zhang}
\author[1]{P. J. Birrell}
\author[2]{N. Boddington}
\author[1]{A. M. Presanis}
\author[2]{R. G. Pebody}
\author[1,2]{D. De Angelis}
\affil[1]{{\small Medical Research Council, Biostatistics Unit - University of Cambridge, School of Clinical Medicine}}
\affil[2]{{\small Centre for Infectious Disease Surveillance and Control, Public Health England}}
\affil[$ ^\dagger $]{{\small Author for correspondence: \href{mailto:alice.corbella@mrc-bsu.cam.ac.uk}{alice.corbella@mrc-bsu.cam.ac.uk}}}
\affil[$ ^\S $]{{\small Equal contributor}}
\maketitle

\begin{abstract}
\textbf{Background}
Influenza remains a significant burden on health systems. Effective responses rely on the timely understanding of the magnitude and the evolution of an outbreak. For monitoring purposes, data on severe cases of influenza in England are reported weekly to Public Health England. These data are both readily available and have the potential to provide valuable information to estimate and predict the key transmission features of seasonal and pandemic influenza.

\textbf{Methods} 
We propose an epidemic model that links the underlying unobserved influenza transmission process to data on severe influenza cases. Within a Bayesian framework, we infer retrospectively the parameters of the epidemic model for each seasonal outbreak from 2012 to 2015, including: the effective reproduction number; the initial susceptibility; the probability of admission to intensive care given infection; and the effect of school closure on transmission. The model is also implemented in real time to assess whether early forecasting of the number of admission to intensive care is possible.

\textbf{Results}
Our model of admissions data allows reconstruction of the underlying transmission dynamics revealing: increased transmission during the season 2013/14 and a noticeable effect of Christmas school holiday on disease spread during season 2012/13 and 2014/15. When information on the initial immunity of the population is available, forecasts of the number of admissions to intensive care can be substantially improved.

\textbf{Conclusion} 
Readily available severe case data can be effectively used to estimate epidemiological characteristics and to predict the evolution of an epidemic, crucially allowing real-time monitoring of the transmission and severity of the outbreak.

\end{abstract}

\section*{Background}

Recent annual epidemics of influenza have resulted in about 3 to 5 million cases of severe illness each season worldwide \cite{whoFactsheet}. Historically, influenza has always placed a large burden on many national health systems \cite{Pitman2007}, particularly as a result of severe cases in the most at risk groups \cite{Hayward2014} (e.g. elderly \cite{matias2014estimates}, children and people with underlying chronic medical conditions \cite{neuzil2000burden}, persons living in deprived areas\cite{zhao2015ethnicity}, etc. ). 

Measures of different characteristics of an outbreak, whether from seasonal or a newly emergent strain, 
are crucial to understand the healthcare burden and plan appropriate response measures. For seasonal influenza, retrospective knowledge of severity and transmissibility provides a much valuable baseline measure against which to compare the severity and transmissibility of future pandemics. Prospectively, predictions of the likely extent of transmission and the resulting number of severe cases are crucial to anticipate demands on health care facilities (e.g. number of beds in hospital) for each season. These timely predictions are even more crucial to inform prompt targeted response in the event of a new emerging strain with the potential to cause a pandemic \cite{Kerkhove2010}. 

Epidemic models are increasingly used to understand the effect of particular interventions including: vaccination policies \cite{Baguelin2013}; school closures to reduce transmission in a pandemic \cite{House2011, TeBeest2015, Vynnycky2008}; reinforced use of antiviral drugs \cite{Ferguson2006}; or changes in hospital management policies.

These models are generally applied to data, such as \ac{GP} consultations for \ac{ILI} \cite{Birrell2011, Baguelin2013} or health-related online queries \cite{Yang2015} which are only loosely related to the actual burden and are characterized by highly volatile noise.

By contrast, more specific timely data on a sample of confirmed cases (e.g. confirmed influenza hospitalizations) might be collected routinely by national health systems. An example of these data is the \ac{USISS} \cite{EngSurv} that records counts of the weekly \ac{ICU} and \ac{HDU} admissions and deaths with confirmed influenza in all hospital trusts in England.

Only recently, and in the context of a pandemic, has some attention been paid to estimating and predicting pandemic transmission from routinely collected confirmed-case data \cite{Shubin2016}. This has entailed the development of a very complicated model which is difficult to use in a seasonal monitoring setting (when less effort is placed on data collection) with a prediction goal.
Here we explore a much simpler model to be applied to seasonal influenza, and possibly during a pandemic, relying only on simpler data on severe cases alone, which are timely available.
We therefore investigate if data collected through \ac{USISS} can characterise both seasonal and pandemic epidemics, aiming to achieve both the estimation and the prediction goal.

We formulate an epidemic model that links the available USISS data to the underlying unobserved dynamics of influenza in the UK. The model parameters are inferred using data from the seasonal epidemics in 2012-2015, to obtain nation-level estimates of transmission, as measured by $R_n$, the average number of new cases generated by an infectious individual in a partially immune population, and severity, as measured by the probability of \ac{ICU} admission given infection. 

Additionally, to assess the predictive power of the model, we perform analyses at different dates within each season. Finally, we study what would happen in the event of a pandemic, when the USISS surveillance scheme would be upgraded to collect more information.
\section*{Methods}

\subsection*{Data}

Following the 2009 pandemic, the \ac{WHO} declared the beginning of a post-pandemic phase \cite{who2}, encouraging national public health agencies to establish hospital-based surveillance systems to monitor the epidemiology of severe influenza. 
In response to these guidelines, and to understand the baseline epidemiology of severe influenza, the UK developed a surveillance system to monitor severe cases of influenza, the \ac{USISS} \cite{prot1, prot2}. 
After a pilot phase in 2010/11, USISS has run for each influenza season, providing data on laboratory-confirmed \ac{ICU}/\ac{HDU} influenza cases and on laboratory-confirmed hospitalized cases.

According to the USISS protocol \cite{prot1}, all \ac{NHS} trusts report the weekly number of laboratory-confirmed influenza cases admitted to \ac{ICU}/\ac{HDU} and the number of confirmed influenza deaths in \ac{ICU}/\ac{HDU} via a web tool. An \ac{ICU}/\ac{HDU} case is defined as a person who is admitted to \ac{ICU}/\ac{HDU} and has a laboratory-confirmed influenza A (including H1, H3 or novel) or B infection. 

USISS runs annually from week 40 to week 20 of the following year but, in the event of a pandemic, it can be activated out of this window and will collect the same data at all levels of care, not only \ac{ICU}/\ac{HDU}.

\begin{figure}[h]
\captionsetup{width=0.96\textwidth}
\centering
\includegraphics[scale=0.65]{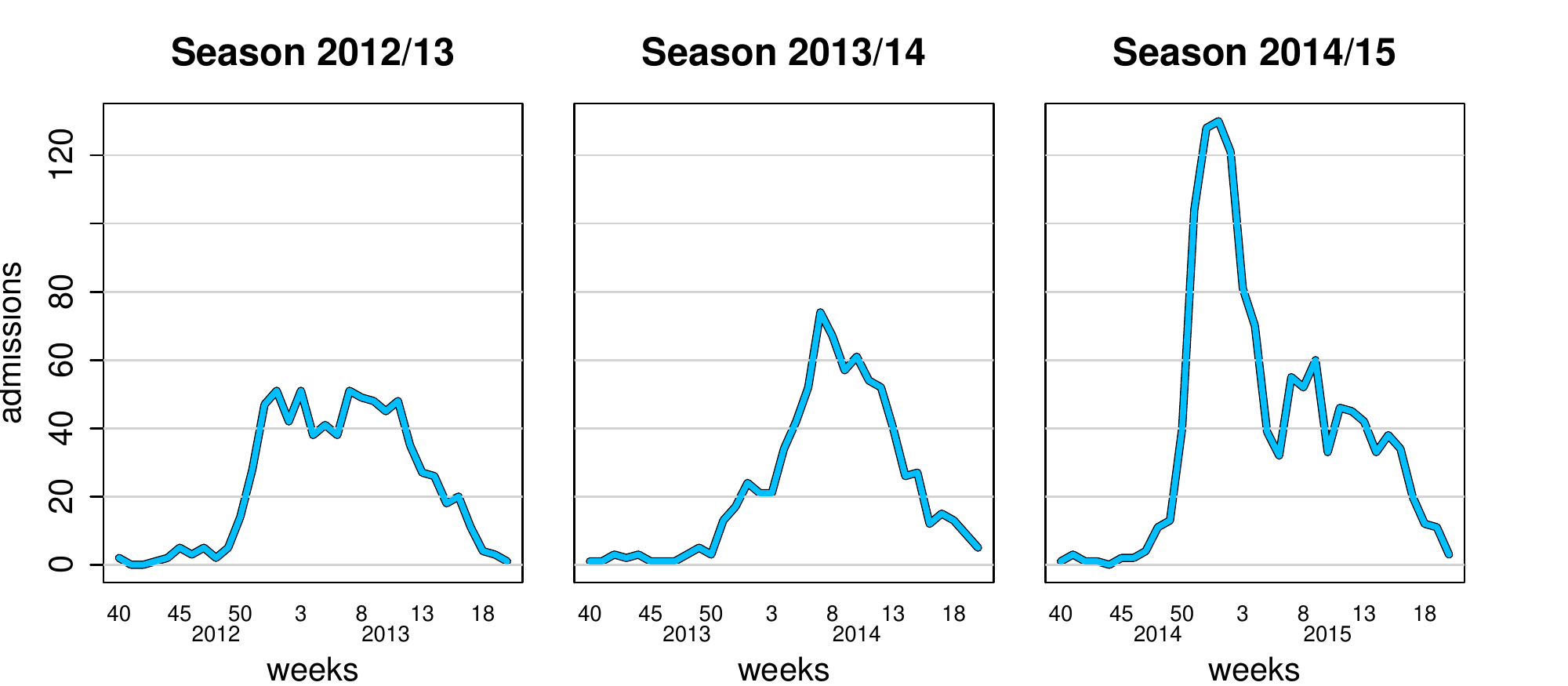}
\caption{\textbf{Weekly \ac{ICU}/\ac{HDU} admissions by season.} Time is measured in week number as reported on the x axis.}
\label{f1}
\end{figure}

Data are broken down by age group and influenza type/subtype. Total \ac{ICU}/\ac{HDU} admissions between 2012 and 2015 are shown in Figure \ref{f1}, varying substantially across seasons. In the 2012/13 season, mainly characterized by Influenza B and Influenza A(H3N2) outbreaks, the number of admissions peaks early, maintaining this plateau for several months \cite{ANNrep1213}. In 2013/14, when the predominant strain was A(H1N1), the time series displays a smoother increase, a well localized peak and a subsequent regular decrease \cite{ANNrep1314}. Lastly, in 2014/15, the number of \ac{ICU} admissions peaks earlier and has a dramatic drop at the beginning of the new year, which is followed by a smaller wave resulting in a time series characterized by a double peak. During this season, Influenza A(H3N2) was the predominant virus circulating and the total number of \ac{ICU} admissions was higher; this strain is well-known to lead to more severe outcomes, particularly in the elderly \cite{ANNrep1415}. 

\subsubsection*{Additional sources of information}

In addition to the mandatory scheme, a subgroup of NHS trusts in England is recruited every year to participate in the \ac{USISS} sentinel scheme \cite{prot2, BODDINGTON2017}, which reports weekly number of laboratory-confirmed influenza cases hospitalised at all levels of care. From this scheme, individual-level data on all \ac{ICU}/\ac{HDU} admissions (until season 2012/13) or on hospital admissions in the young ($ \leq 17 $ years old) population (from season 2013/14 onwards) are available, including clinical details such as date of symptom onset, of hospital and \ac{ICU} admission, and date of discharge from \ac{ICU}.

These data provide useful information on the process between influenza infection and \ac{ICU} admission (e.g. the time elapsing from symptom onset to ICU admission). Further information on this process (e.g. proportion of symptomatic cases) can be found in the existing literature about the incubation period of influenza \cite{Tom2011} and the hospitalization fatality rate \cite{Presanis2014}.

\subsection*{Model}

\begin{figure}[h!]
\centering
\captionsetup{width=0.96\textwidth}
\includegraphics[scale=1]{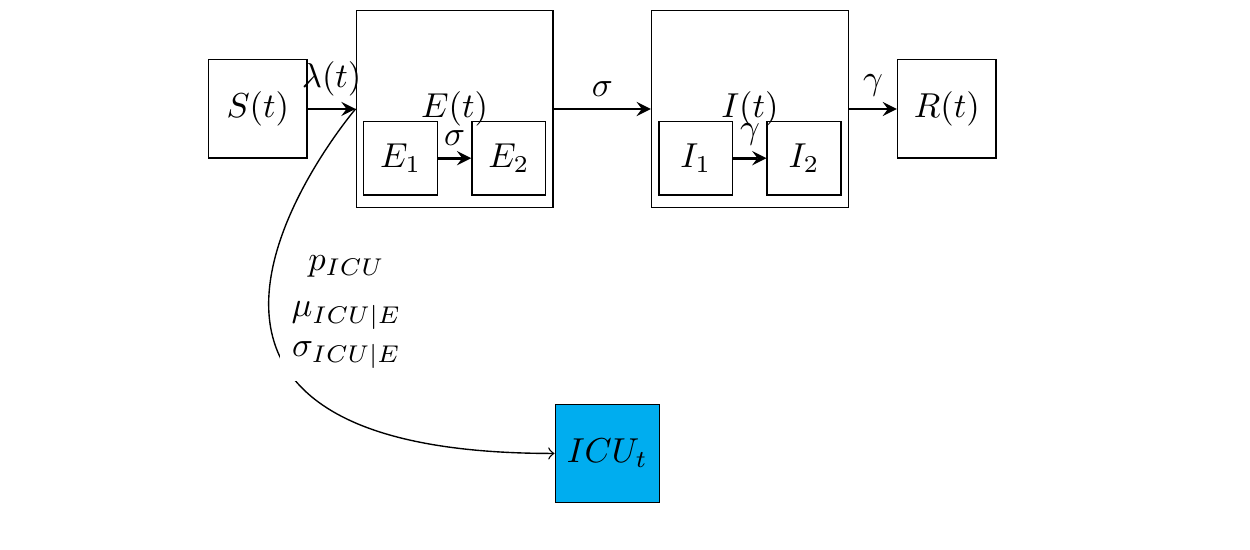}
\caption{Schematic diagram representing the epidemic model and the model linking transmission to  \ac{ICU}/\ac{HDU}  admissions (in blue).}
\label{f2}
\end{figure}

We used an epidemic model (Figure \ref{f2}) to describe the spread of influenza in England \cite{keeling2008modeling}.
We assumed that the population changes according to a deterministic model in continuous time. Time is measured in days and denoted by $ t $. 

The population is divided according to health status into four compartments: susceptible ($ S $), exposed ($ E $), infectious ($ I $) and removed ($ R $). The $ E $ and $ I $ compartment are further divided into two ($ E_1, E_2$ and $ I_1, I_2, $ respectively) leading to waiting times in the $ E $ and $ I $ states, distributed according to gamma distributions \cite{Wearing2005}. In the formulas below, the letters $ S,E_1, E_2, I_1, I_2, R $ denote the number of people in each compartment. The total size of the population is fixed over every season and denoted by $ N $. 
The change of compartment is determined by the transition rates: $ \lambda (t) $, $ \sigma $ and $ \gamma $ explained below. 

The infection rate $ \lambda(t) $ is proportional to the proportion of people in the infectious compartment at $ t $, $\frac{I_1(t)+I_2(t)}{N} $ and a piecewise constant transmission rate $ \beta^* (t)$ (the rate at which new infections take place):

\begin{equation} \label{e1}
\lambda(t) = \beta^*(t) \frac{I_1(t)+I_2(t)}{N}.
\end{equation}
$ \beta^*(t) $ is piecewise constant and it allows for a scaling factor $ \kappa  \in (0,1]$ that expresses the change of the base contact rate $ \beta $ due to school closure \cite{TeBeest2015} as reported in Equation \ref{e2}.

\begin{equation}\label{e2}
\beta^*(t) = 
\begin{cases} \kappa\cdot \beta, & t \in \mbox{school holidays} \\ 
\beta, & \mbox{otherwise}. 
\end{cases}
\end{equation}

The transition rates $ \sigma $ and $\gamma $ are related to the mean latent period, $ d_L $, and the mean infectious period, $ d_I $, by:

\begin{equation}\label{e3}
\sigma = 2/d_L , \text{\hspace*{3cm}}  \gamma = 2/d_I
\end{equation}

The system of differential equations that defines the epidemic model is reported in Equation \ref{e4}.

\begin{equation}
\begin{split}
\frac{d S}{d t} &=  -\lambda(t) \cdot S\\
\frac{d E_1}{d t}&= \lambda(t) \cdot S - \sigma \cdot E_1\\
\frac{d E_2}{d t}&= \sigma \cdot E_1 - \sigma \cdot E_2\\
\frac{d I_1}{d t}&= \sigma \cdot E_2 - \lambda \cdot I_1\\
\frac{d I_2}{d t}&= \lambda \cdot I_1 - \lambda \cdot I_2\\
\frac{d R}{d t}&= \lambda \cdot I_2\\
\end{split}
\label{e4}
\end{equation} 
Here we have assumed homogeneous mixing among contacts (i.e. people are all equally likely to meet, irrespective of their age class and residence, for example). 

This transmission model is linked to the data on \ac{ICU} admissions through an observational model that defines the time elapsing from infection to \ac{ICU} admission and the probability of \ac{ICU} admission conditional on infection.  

Denote with $ f_{ICU|I}(w) $ the probability that $ w $ weeks elapse from infection to \ac{ICU} admission, and with $ p_{ICU} $ the probability of \ac{ICU} admission given infection. We can link $ \mu_w $, the average number of \ac{ICU} admissions during week $ w $, to the weekly new infections in the previous weeks via a convolution:
\begin{equation}\label{eq5}
\mu_w  =\sum_{v=0}^{w}f_{ICU|I}(w-v)\cdot \Delta I_v  p_{ICU} 
\end{equation}
where $ \Delta I_v=(S(7v-7)-S(7v))$ is the count of the new infections during week $ v $.

To formulate the likelihood of the data, we assumed that the observed number of \ac{ICU} admissions is the realisation of a Negative Binomial random variable centred on $ \mu_w $ with over dispersion parameter $ \eta $:
\begin{equation}\label{eq6}
ICU_w \sim \text{NegBin} (\mu_w, \eta),
\end{equation}
i.e $ ICU_w $ has density function:
\begin{equation}\label{eq7}
f(ICU_w=x) = \frac{\Gamma(x+r_w)}{\Gamma(x)\Gamma(x+r_w)}\left(\frac{1}{\eta}\right)^{r_w} \left(1-\frac{1}{\eta}\right)^x
\end{equation}
with $r_w=\frac{\mu_w}{\eta-1}$. 

The Additional file  contains the full specification of the transmission model, its re-parametrization and full derivation of  $ f_{ICU|I}(w) $.

\subsection*{Parameter estimation}
To define the epidemic we need to estimate or set both the transitions rates parameters  (i.e. $ \beta, \kappa, \sigma, \gamma $) and the initial state of the epidemic (i.e. $ S(0), E_1(0), E_2(0), I_1(0), I_2(0), R(0) $). 

The epidemic model can be re-parametrized \cite{Wearing2005} and a number of quantities may be defined, including: $ \pi $, the initial proportion of non-immune people; $ I^{tot}(0)=(I_1(0)+I_2(0)) $, the total number of infectious people at $ t=0 $; the basic reproduction number $R_0$ that is the average number of successful transmissions per infectious person in a fully susceptible population; and the effective reproduction number $R_n$ that is the average number of successful transmissions per infectious person in a partially susceptible population. All these parameters are useful under a health-policy perspective. 

The parameters $\sigma$ and $\gamma$ are assumed known from previous studies \cite{Tom2011, Birrell2011}, as they can be inferred only with detailed information at the individual level. Likewise, the population size $N $ is assumed known and fixed to the values estimated by the \ac{ONS} \cite{ons}. 

We used a Bayesian approach to draw inference on the other parameters. Bayesian inference consists in summarizing prior information on a general parameter $ \theta $ in a distribution $ \pi(\theta) $ and updating it with the information deriving from a set of data $ x $, contained in its likelihood $ \mathcal{L} (\theta|x) $, to derive the posterior distribution: 
\begin{equation}\label{eq8}
p(\theta|x) \propto \pi(\theta) \cdot \mathcal{L} (\theta|x).
\end{equation}

We considered two scenarios. In the first one we assumed we have no prior information on the values of the parameters except for lower and upper bounds, hence the prior distributions on all the parameters are non-informative (see Additional file 1). Table \ref{t1} lists the lower and upper limits of some transformations of the parameters and the values assumed known in this scenario.

\begin{table}[h!]
\centering
\caption{Prior distributions of the parameters in the non-informative scenario}
\begin{tabular}{cccc}
\hline
Unknown parameters definition & Parameter & Lower limit & Upper limit\\ 
\hline
Susceptibility &   $\pi$ & 0 & 1\\ 
Initial number of infectious &  $I^{tot}(0)=(I_1(0)+I_2(0))$ &  0 & 10000 \\ 
Transmission rate &  $\beta$ &  0 & 1.12 \\ 
Over-dispersion &  $\eta$ & 1 & 100 \\ 
P of ICU admission given infection &  $p_{ICU}$ & 0 & 1 \\ 
Scaling factor for school closure &  $\kappa$ &  0 & 1 \\ 
\hline
Parameters assumed known \hspace*{0.4cm} &  \hspace*{0.2cm}  Parameter \hspace*{.2cm}&  \hspace*{.2cm}Value  \hspace*{.2cm}&\\ 
\hline
Rate of becoming infectious &  $\sigma$ &  1&\\
Rate of recovery &  $\gamma $ & 0.5797 & \\  
Population of 2012/13  &  $N_{\text{2012/13}}$ & 53,679,750& \\
Population of 2013/14  &  $N_{\text{2013/14}}$ & 54,091,200 &\\
Population of 2014/15  &  $N_{\text{2014/15}}$ & 54,551,450 &\\
\hline
\end{tabular}
\label{t1}
\end{table}

In the second scenario we used sero-prevalence data from the 2010/11 season \cite{Hoschler2012} to formulate a prior distribution for the initial susceptibility $\pi$. The use of sero-prevalence data to describe the immunity of a population could be debatable, since the results may be extendible only to seasons with similar predominant strains circulating. Here, sero-samples were taken during an H1 predominant season: this sub-type was prevalent also in the 2012/13 season, but not in 2014/15. However, combining this prior with the data allows us to test how much prior knowledge is needed to overcome the lack of information about susceptibility from the data. We also derived an informative prior distribution on $ p_{ICU} $ by combining estimates of the probability of hospitalization given infection from a previous severity study \cite{Presanis2014} with estimates of the probability of ICU/HDU admission given hospitalization from the aggregate data of the USISS sentinel scheme. 
Table \ref{t2} lists the prior distributions of the two parameters that change in the informative scenario. The remaining parameters are again assumed to be uniformly distributed.

\begin{table}[h!]
\centering
\caption{Prior distributions of the parameters that change in the informative scenario}
\begin{tabular}{cc} 
\hline
 Parameter & Distribution \\ 
 \hline
  $\pi$ & $\sim  \text{LogNorm} (\log \mu=\log( 0.401), \log \sigma = 0.2)$ \cite{Hoschler2012}\\ 
 $p_{ICU}$ & $\sim \text{LogNorm} (\log \mu=\log (0.000239), \log \sigma = 1)$ \cite{Presanis2014}\\ 
 \hline
\end{tabular}
\label{t2}
\end{table}

\subsection*{Analyses}

\begin{figure}[h!]
\centering
\captionsetup{width=0.96\textwidth}
\includegraphics[scale=0.45]{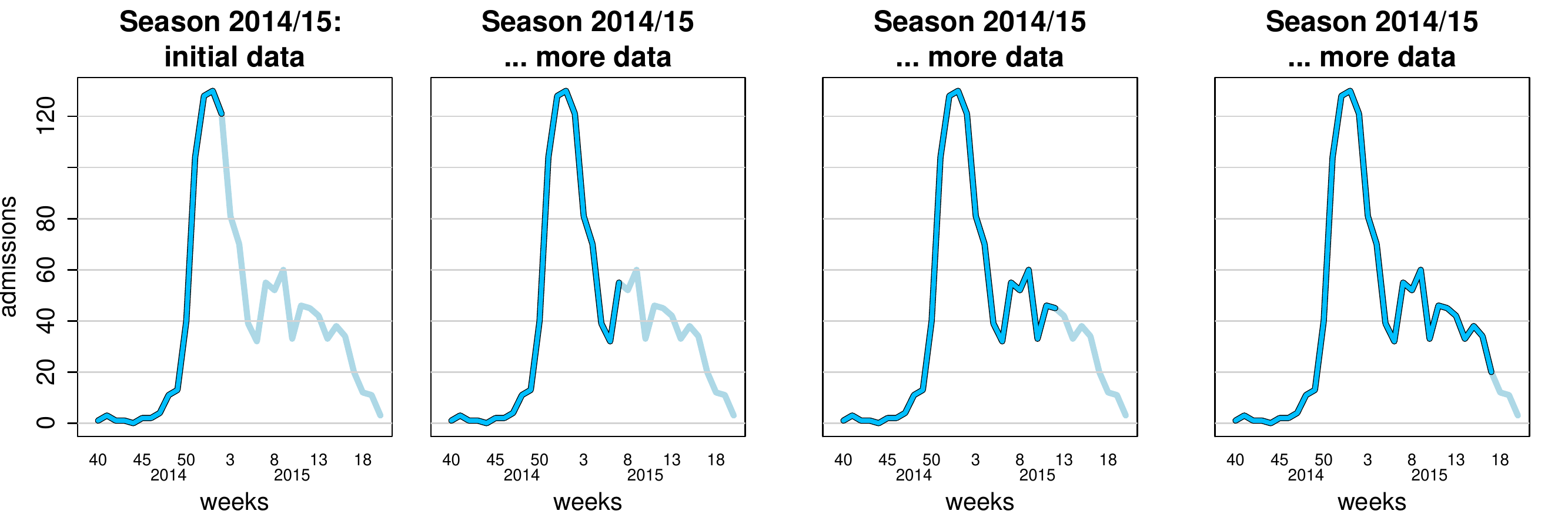}
\caption{\textbf{Prospective analysis.} Sequential data on \ac{ICU} admissions used in the prospective analysis in the season 2014/15.}
\label{f3}
\end{figure}

For both the prior settings we performed two types of analysis: firstly we considered all the data reported in Figure \ref{f1} and we analysed them retrospectively. Secondly, to assess the predictive ability of our model, we performed estimation and forecasting assuming only an initial portion of the data are available. We used the data up to week $ w $ as a training dataset to estimate the parameters. Then we predicted the evolution of the epidemic after week $ w $, based on the estimates from the training dataset.
We tested the following prediction time points: $ w= 3,8,13 $, and 18 from the beginning of the new year. An example of the sequence of data used to analyse prospectively an epidemic (season 2014/15) is reported in Figure \ref{f3}.

To approximate the posterior distribution, we used a Metropolis Hastings block updated sampling algorithm \cite{robert2009introducing}, coded using the \texttt{R} programming language \cite{R}. The system of differential equations \ref{e4} was solved using the \texttt{R} package \texttt{deSolve} \cite{soetaert2010solving}. Details on the algorithm are available in Additional file 1 and the code is available at \url{http://www.mrc-bsu.cam.ac.uk/software/miscellaneous-software/}.

\section*{Results}

\subsection*{Retrospective analysis}
\begin{figure}[h!]
\centering
\captionsetup{width=0.96\textwidth}
Season 2012/13 \hspace{2cm} Season 2013/14 \hspace{2cm} Season 2014/15\\
\includegraphics[trim={0 7cm 0 0}, scale=.6]{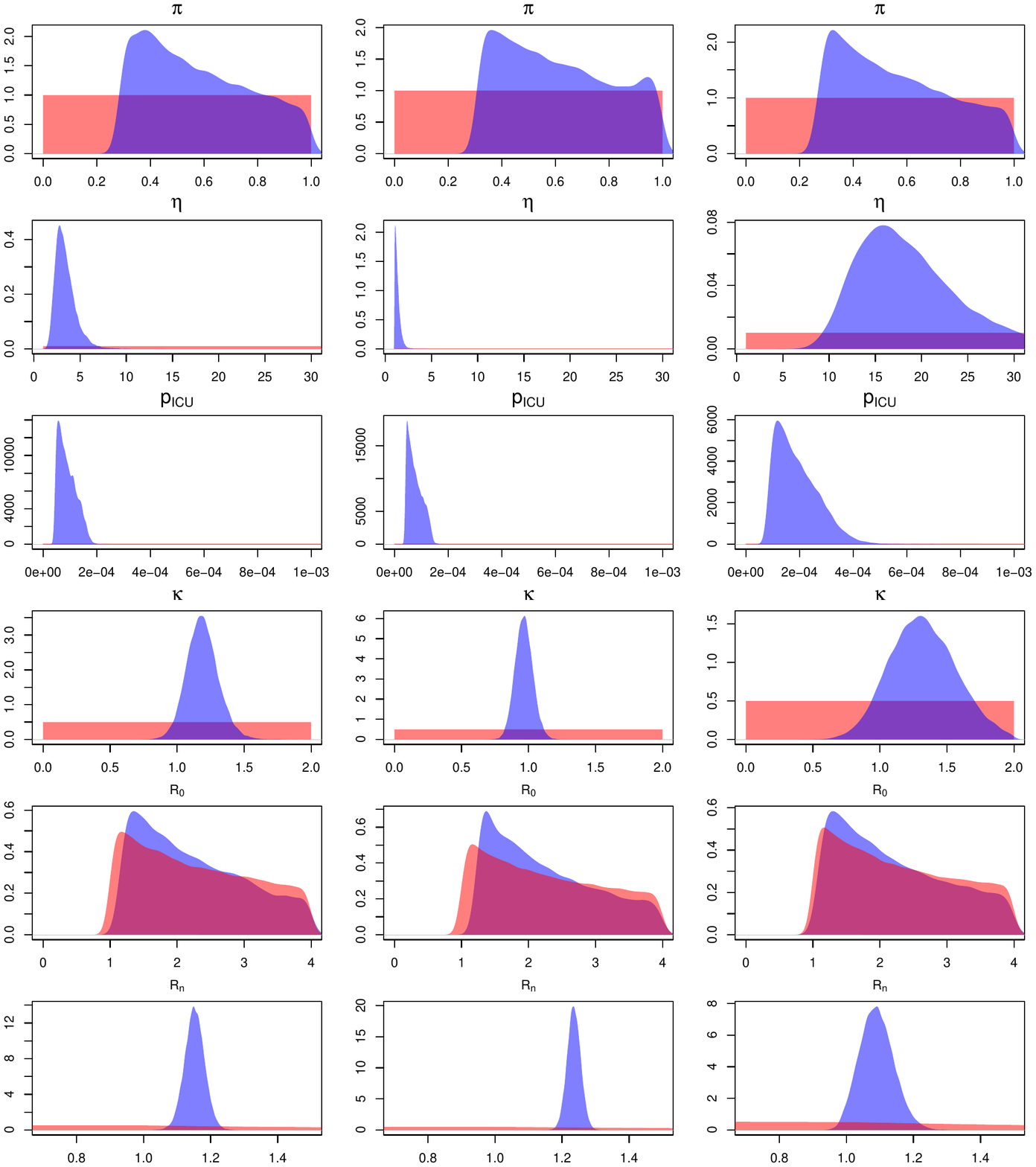}
\caption{\textbf{Retrospective analysis.}
Prior (red) and posterior (blue) distributions of: the initial susceptibility ($ \pi $); the over-dispersion parameter ($ \eta $); the probability of ICU admission given infection ($ p_{ICU} $); the scaling parameter ($ \kappa $); and the basic and effective reproduction number ($ R_0$ and $ R_n $). The results are derived from season 2012/13 (left column), season 2013/14 (centre) and season 2014/15 (right column). }
\label{f4}
\end{figure}

The retrospective analysis of the data was first performed in the uninformative scenario. The resulting posterior distributions are displayed in Figure \ref{f4} with the posterior median and 95\% \ac{CrI}s of some of the parameters reported in Table \ref{t3}. Note that the posterior distribution of the initial susceptibility $ \pi $ and the basic reproduction number $ R_0 $ are almost identical to the prior. This is due to the fact that the information contained in the data is not sufficient to determine separately the values of the parameters describing both the initial immunity and the transmission rate. This problem is explored in detail in Additional file 1.

Data are much more informative about parameters $ \eta $, $ p_{ICU}$ and $ \kappa $. The highly variable behaviour of the ICU admissions count in season 2014/15 is reflected by the over-dispersion parameter $\eta $, whose distribution is significantly higher compared to the ones estimated from the 2012/13 and 2013/14 seasons. The range of the probability of going to ICU given infection, $ p_{ICU} $, is always between 0.04\% and 0.4\%. Its median is higher in season 2014/15, in agreement with the higher severity that was detected during this influenza season \cite{BODDINGTON2017}. The multiplicative factor $ \kappa $ introduced to allow for a school-closure effect is centred on 1 for season 2013/14 and centred around higher values in the remaining seasons. A possible explanation for this counter-intuitive phenomenon relies on the age distribution of the sample population. Our data have a different distribution compared to the English population (\cite{BODDINGTON2017}, \cite{ons}), with patients over 65 being over represented and children in school years being under represented.  The elderly individual perhaps are more likely to meet other potential influenza spreaders (e.g. children) during school closures, particularly over Christmas holiday. It makes sense, therefore, to observe an inverse relationship between school closure and the transmission rate, in contrast to results that might be expected from a more representative sample of the population \cite{TeBeest2015}.
However, this piecewise increment in transmission rate may incorporate other time-varying phenomena that affect the force of infection. The Christmas holiday often coincides with the beginning of a colder and more humid period and changes in vapour pressure, that might imply an increasing spread of influenza \cite{Lipsitch2009a}. 
Lastly the posterior median of the effective reproduction number $ R_n $ is equal to 1.152, 1.235, 1.089 in seasons 2012/13, 2013/14 and 2014/15 respectively.

\begin{table}[ht]
\caption{Posterior medians and 95\% \ac{CrI}s from the retrospective analysis of the ICU admissions with uninformative priors.}
\label{t3}
\centering
\begin{tabular}{rlll}
  \hline
 & 2012/13 & 2013/14 & 2014/15 \\ 
  \hline
$ \pi $ & 0.546 (0.297 - 0.969) & 0.589 (0.32 - 0.977) & 0.531 (0.28 - 0.968) \\ 
 $I^{tot}$ & 4106 (1441 - 11510) & 1357 (484 - 3312) & 9590 (3053 - 28493) \\ 
   $\beta$ & 0.611 (0.344 - 1.126) & 0.608 (0.367 - 1.118) & 0.596 (0.324 - 1.119) \\ 
  $\eta$ & 3.204 (1.888 - 6.101) & 1.25 (1.011 - 2.096) & 17.925 (10.412 - 35.812) \\ 
  $p_{ICU}(\cdot10^2)$ & 0.084 (0.046 - 0.161) & 0.071 (0.042 - 0.134) & 0.175 (0.085 - 0.374) \\ 
 $ \kappa $ & 1.185 (0.971 - 1.434) & 0.965 (0.841 - 1.1) & 1.313 (0.866 - 1.824) \\ 
 
  $ R_n $ & 1.152 (1.093 - 1.211) & 1.235 (1.196 - 1.275)  & 1.089 (0.997 - 1.195)   \\
   \hline
\end{tabular}
\end{table}

\begin{figure}[h!]
\centering
\captionsetup{width=0.96\textwidth}
\includegraphics[trim={0 11.5cm 0 0},scale=0.42]{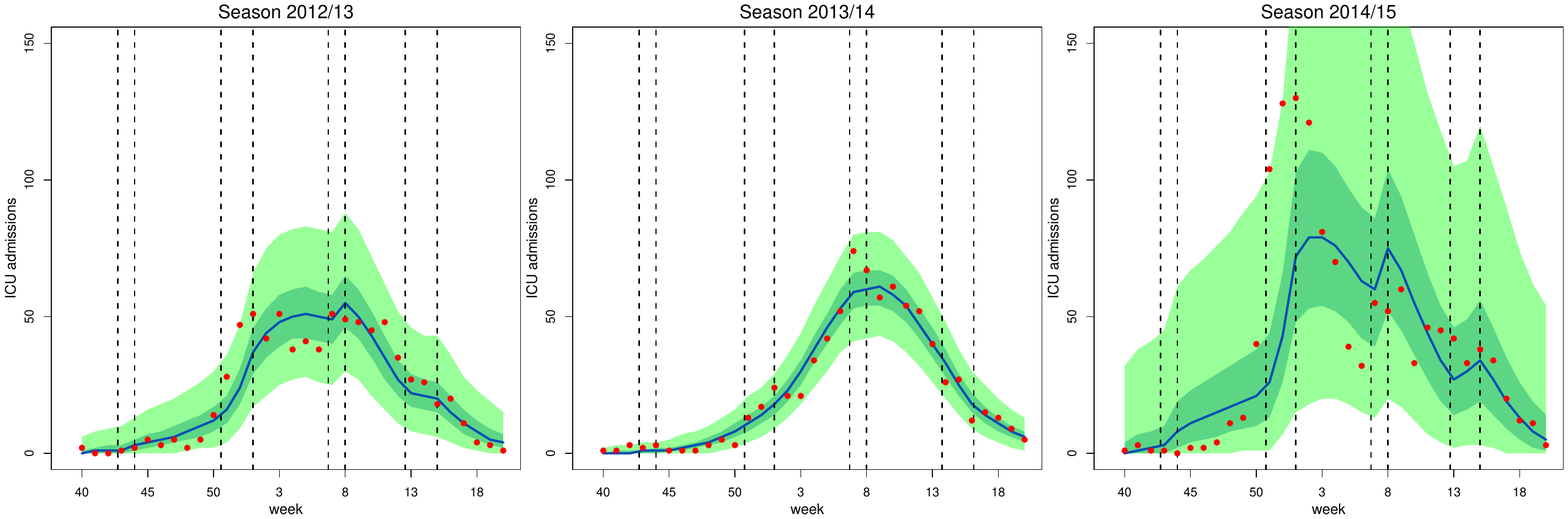}
\caption{\textbf{Retrospective analysis.}
Median (blue), 95 \% \ac{CrI} (light green) and quartile (dark
green) of the posterior predictive distributions and observed values (red) for the weekly ICU/HDU admissions across seasons. The vertical dashed lines represent the breakpoints for the piecewise transmissibility $ \beta^*(t) $ (i.e. start and end of each school holiday).}
\label{f5}
\end{figure}

Although the \ac{CrI}s of the parameter $ \kappa $ included 1, the posterior probability of it being larger than 1 ($\text{Pr}( \kappa > 1) $) is substantial for two season. The introduction of this parameter allows the flexibility needed to represent the specific features of each season. This can be observed in the posterior predictive distribution of the weekly ICU admissions reported in Figure \ref{f5}. Specifically in season 2012/13 we manage to reproduce the plateau that takes place from the end of the Christmas vacations to the February half term. Regarding instead the double peaking season of 2014/15, the 95\% Credible bounds are not narrow, but the timing of the peak of the distribution is predicted substantially better than in the case of constant infection rate (results not shown).

The same analysis was performed in the second scenario, i.e. allowing informative priors on the susceptibility $ \pi $ and on $ p_{ICU} $ as defined in Table \ref{t2}.
The introduction of these prior distributions compensate for the lack of information, allowing the identification of $ \pi $ and improving the precision of the posterior distribution of $ p_{ICU} $. This affects also other parameters such as $ \beta $ and $ R_0 $. However, their the posterior distributions are driven by the prior distributions alone, and they do not learn from the data. In terms of fit there was no improvement. Results are reported in Additional file 1.

\subsection*{Prediction}
\begin{figure}[h!]
\centering
\captionsetup{width=0.96\textwidth}
Season 2012/13 \hspace{2cm} Season 2013/14 \hspace{2cm} Season 2014/15\\
\includegraphics[trim={0 9cm 0 0}, scale=.6]{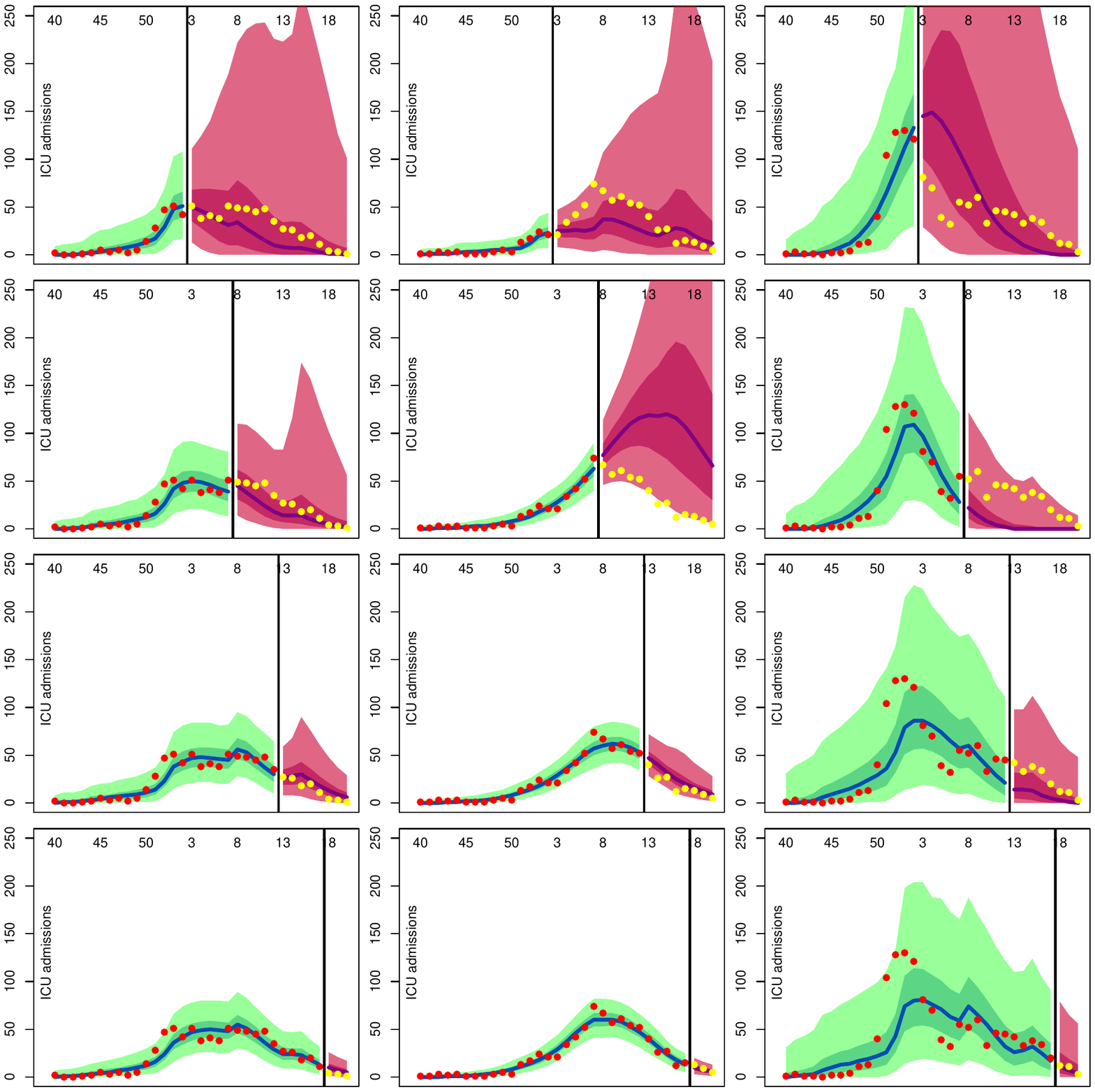}
\caption{\textbf{Prospective analysis.}
The black line displays the analysis time; the blue line and green shaded area represent median, quartile (dark green) and 95\% \ac{CrI}s (light green) of the posterior predictive distribution for the training dataset weeks. The pink area displays posterior quartiles (deep pink) and 95\% \ac{CrI}s (light pink) for the predicted future observations, and the purple line displays the median; the red dots are the training data and the yellow dots are the observations we have predicted.}
\label{f6}
\end{figure}

The prospective analysis of the data in the uninformative scenario resulted in very wide predictions of the future dynamics, therefore we assumed the informative priors reported in Table \ref{t2}.
The performance of the model at different times is plotted in Figure \ref{f6} for each season. 

Season 2013/14, despite displaying the most regular data, is the most difficult to predict: the well-defined initial growth biases the predictions towards a major outbreak. This leads to overestimation of the median and the credible intervals of the posterior predictive distribution until mid-march (week 13 from the beginning of the year). For the other two seasons, the median predicted weekly ICU admissions is always very close to the data points, but the credible intervals narrow to reasonable bounds only towards the end of February (week 8 from the beginning of the year).

In spite of the simplicity of our model, the flexibility introduced by the parameter $ \kappa $ allows for the correction ``on the fly" of the prediction, adapting to new peaks (e.g. season 2014/15) or periods of constant influenza circulation (e.g. season 2012/13).

\subsection*{Further results}
We simulated the weekly count of Hospital admissions in the case of a pandemic and we extended our model enabling the inference of the parameters from these data. Despite the increased number of observations, the model performed very similarly to the case of non-pandemic \ac{ICU}-counts data. We diagnosed identifiability problems in the uniform prior scenario and predictions were good only when more informative prior distributions (on the susceptibility and probability of hospitalization) was included. Results from this analysis are reported in Section 5 of Additional file 1.

Other analyses performed include: prospective analysis for the uninformative scenario and retrospective analysis within the informative scenario. Results of these analysis are reported in Section 4 of Additional file 1.

\section*{Discussion}
In this paper we proposed a model to estimate and predict influenza outbreaks from routinely collected data on admissions to \ac{ICU}/\ac{HDU}. 

We investigated the performance of the proposed model both on simulated and on real data. By fitting the model to simulated number of weekly ICU admissions, we discovered that, even with very vague prior information, we could obtain estimates of some of the main parameters, including the initial infection rate, the probability of going to ICU given infection, the effective reproduction number $ R_n $ and the scaling factor for school holidays $ \kappa $.
When we injected information on the distribution of the average immunity ($ 1-\pi $) and on $ p_{ICU} $, estimates of the remaining parameters could be obtained. We were also able to forecast the evolution of the outbreak by analysing the first months of the epidemic using data up to the peak of influenza activity.

The model was applied to real data on the weekly number of ICU admissions from seasons 2012/13, 2013/14 and 2014/15, confirming the performance obtained on the simulated data. The estimated values of the effective reproduction number $R_n$ were similar to those estimated during the past decade of seasonal influenza \cite{Baguelin2013}. A scaling parameter allowed the transmission rate to vary between school and holiday/half-term periods,  which resulted in a good fit of the  model to the data.

Recently, a similar analysis was performed on the Finnish influenza pandemic of 2009 \cite{Shubin2016} using a more elaborate model, analysing confirmed data on both hospitalizations and \ac{GP} consultation. Their inclusion of GP data enhances the performance of the inference. Nevertheless, these data are harder to collect in a larger population (England is almost 10 times more populated than Finland) and out of pandemic emergencies. By contrast, the inference performed through our model is driven by few data, though readily available, even in real time, in seasonal settings. 
A further advance of the model by \cite{Shubin2016} is that the transmission parameter is time varying according to a Gaussian Process: this allows an accurate description of the past dynamics but makes the prediction hard, since this temporal variation cannot be forecast. By contrast, our simple piecewise constant model is able to well forecast the future trend and it includes enough flexibility to describe appropriately the present and the past data. 

Our work has also some limitations: firstly, our model is non-age-specific. This was dictated by the very small data size which did not allow sub-grouping. Secondly, the quality of some estimates and predictions strongly relies on prior information on the proportion of non-immune people. As this information is needed to overcome the lack of identifiability in the parameters, we used sero-prevalence data following the 2010/11 epidemic. This is not likely to be correct for all the three seasons analysed, as the predominant strain circulating was different across seasons. Likewise, the model that describes the time elapsing between infection and \ac{ICU} admission, is assumed to be fixed and mostly known, but this assumption is not likely to be valid. The other element that defines the observational process, i.e. the probability of \ac{ICU} admission given infection, is also sensible to the choice of prior distribution.

\section*{Conclusion}

The work presented here is a proof of concept of the potential for estimation and prediction of influenza transmission from USISS data. At the same time, the results highlight the need of collecting external data to formulate appropriate prior distribution on the initial immunity of the population, particularly in the event of a pandemic.

The availability of this information, together with the tool we have provided here, allows to retrospectively infer the epidemic parameters from routinely collected data on severe cases during seasonal outbreaks and to predict the temporal dynamics of new epidemics.

\section*{Acronym}
\begin{acronym} 
\setlength{\parskip}{0ex}
 \setlength{\itemsep}{0.5ex}
\acro{USISS}{UK Severe Influenza Surveillance System}
\acro{PHE}{Public Health England}
\acro{ILI}{influenza-like illness}
\acro{GP}{General Practitioner}
\acro{NHS}{National Health Service}
\acro{ICU}{Intensive Care Unit}
\acro{HDU}{High Dependence Unit}
\acro{WHO}{World Health Organization}
\acro{ONS}{Office of National Statistics}
\acro{CrI}{Credible Intervals}
\end{acronym}

\section*{Declarations}

{\small

\subsection*{Availability of data and material}
The datasets used and/or analysed during the current study are available from the author on reasonable request at \href{mailto:richard.pebody@phe.gov.uk}{richard.pebody@phe.gov.uk}.
\subsection*{Competing interests}
 The authors declare that they have no competing interests.
\subsection*{Funding}
AC, DDA, and AMP were supported by the Medical Research Council [grant number MC UI05260666, Program Core SLAH/001]. PJB was supported by the National Institute for Health Research (HTA Project:11/46/03). XSZ, NB, RP, PB and DDA were supported by Public Health England.
\subsection*{Author's contributions}
AC wrote the paper, with help by XSZ DDA and AMP. XSZ and AC wrote the code, with support by PJB and AMP in the formulation of the algorithm. NB provided the data and assisted with their interpretation. DDA and RGP conceived the study. All authors gave a final approval for publication.
\subsection*{Acknowledgements}
The authors would like to thank all the participants of the Armitage Lecture 2015. In particular, Prof. Leonard Held and Dr. Michael Hohle gave substantial advice for the implementation of this model. Likewise, during the Summer Institute on Statistics and Infectious Disease
Modelling \cite{sismid}, Prof. Pejman Rohani helped with suggestions on model evaluation. Finally, AC would like to thank Prof. Rino Bellocco for his support from the very beginning of this project.
}

\bibliographystyle{vancouver} 
\bibliography{reference.bib}  

\end{document}